\documentclass[twocolumn,aps,prl,showpacs,groupedaddress]{revtex4}
\usepackage{mathptmx}
\usepackage[T1]{fontenc}
\usepackage[latin9]{inputenc}
\usepackage{color}
\usepackage{amsmath}
\usepackage{amssymb}
\usepackage{graphicx}
\usepackage{esint}
\usepackage[unicode=true,pdfusetitle,
 bookmarks=true,bookmarksnumbered=false,bookmarksopen=false,
 breaklinks=false,pdfborder={0 0 1},backref=false,colorlinks=true]
 {hyperref}
\hypersetup{
 linkcolor=blue,citecolor=red}

\makeatletter
\@ifundefined{textcolor}{}
{%
 \definecolor{BLACK}{gray}{0}
 \definecolor{WHITE}{gray}{1}
 \definecolor{RED}{rgb}{1,0,0}
 \definecolor{GREEN}{rgb}{0,1,0}
 \definecolor{BLUE}{rgb}{0,0,1}
 \definecolor{CYAN}{cmyk}{1,0,0,0}
 \definecolor{MAGENTA}{cmyk}{0,1,0,0}
 \definecolor{YELLOW}{cmyk}{0,0,1,0}
}

\makeatother

\begin{document}

\title{Protected quasi-locality in quantum systems with long-range interactions }

\author{Lorenzo Cevolani}

\author{Giuseppe Carleo}

\author{Laurent Sanchez-Palencia}

\affiliation{Laboratoire Charles Fabry, Institut d'Optique, CNRS, Univ. Paris
Sud 11, 2 avenue Augustin Fresnel, F-91127 Palaiseau cedex, France}
\begin{abstract}
We study the out-of-equilibrium dynamics of quantum systems with long-range
interactions. Two different models describing, respectively, interacting
lattice bosons and spins are considered. Our study relies on a combined
approach based, on one hand, on accurate many-body numerical calculations
and, on the other hand, on a quasi-particle microscopic theory. For
sufficiently fast decaying long-range potentials, we find that the
quantum speed limit set by the long-range Lieb-Robinson bounds is
never attained and a purely ballistic behavior is found. For slowly
decaying potentials, a radically different scenario is observed. In
the bosonic case, a remarkable local spreading of correlations is
still observed, despite the existence of infinitely fast traveling
excitations in the system. This is in marked contrast with the spin
case, where locality is broken. We finally provide a microscopic justification
of the different regimes observed and of the origin of the protected
locality in the bosonic model.
\end{abstract}


\maketitle
It is common wisdom that the propagation of a signal through a classical
medium presents a distinct notion of causality, characterized by the
progressive time growth of the spatial region explored by the signal.
In spite of the intrinsically non-local nature of the quantum theory,
this familiar notion of locality is preserved in a wide class of quantum
systems with short-range interactions. A milestone example is provided
by the Lieb-Robinson (LR) bounds, which set a ballistic limit to the
propagation of information, with exponentially small leaks outside
the locality cone \cite{lieb1972,Eisert:2015aa}. The existence of
LR bounds has many fundamental implications for thermalization, entanglement
scaling laws, and information transfer in quantum systems \cite{nachtergaele2011}.
A renewed interest in these topics is currently sparked by the impressive
progress in the time-dependent control of ultracold-atom systems.
Direct observation of cone spreading of correlations was reported
in Refs. \cite{cheneau2012,langen2013}.

The extension of the notion of locality to quantum systems with long-range
interactions constitutes a fundamental challenge. The paradigmatic
model of long-range interactions considers an algebraic decay of some
coupling term of the form $V(R)\sim1/R^{\alpha}$ \cite{tagliacozzo2013,manmana2014,pupillo2014,schachenmayer2015,baranov2012}.
It applies either to the exchange coupling term in spin systems, as
realized in cold ion crystals \cite{PhysRevA.72.063407,porras2014},
or to the two-body interactions in particle systems, as realized in
ultracold gases of polar molecules \cite{Buchler:2009aa}, magnetic
atoms \cite{menotti2008}, and Rydberg atoms \cite{pohl2014}. A remarkable
feature of long-range systems is that instantaneous propagation of
information, in violation of locality, can take place when the exponent
$\alpha$ is smaller than some threshold. This possibility is supported
by the known extensions of the LR bounds to long-range interactions
\cite{Hastings2010,Gong2014,FossFeig2015}. The latter yield ``quasi-local''
super-ballistic bounds for $\alpha>d$, where $d$ is the dimension
of the system, whereas for $\alpha<d$ no known generalized bounds
exist, hence suggesting the breaking of quasi-locality. Evidence of
the breaking of quasi-locality in one-dimensional (1D) Ising spin
systems has been reported theoretically \cite{tagliacozzo2013,manmana2014}
and experimentally in cold ion crystals \cite{porras2014,jurcevic2014}.
However, many questions remain open. For instance, although the observations
are compatible with the known long-range bounds, the propagation was
found to be much slower than expected \cite{manmana2014}. Hence,
the bounds are usually not saturated and it is not clear that they
provide a universal criterion for the breaking of quasi-locality.
Moreover the threshold value for the breaking of locality in these
systems is debated, and contrasting results have been put forward
\cite{tagliacozzo2013,manmana2014}. To progress on these questions,
it is of crucial importance to provide a unified understanding of
a wider class of systems and, at the same time, to understand the
microscopic origin of the breaking of quasi-locality.
\begin{figure*}[t]
\includegraphics[width=2\columnwidth]{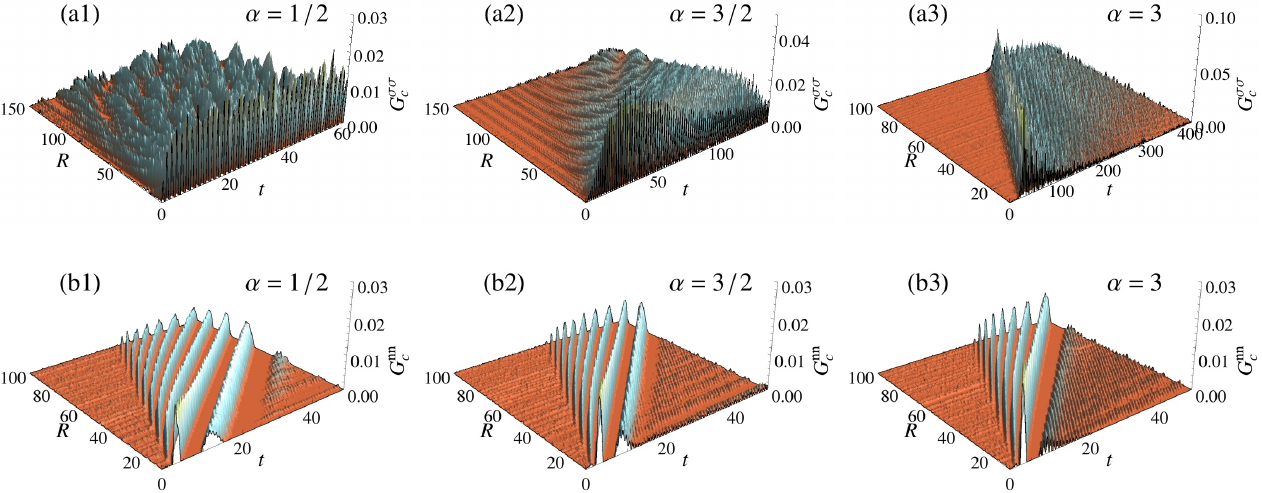}\caption{\textit{\footnotesize \label{fig:tvcm-corr} }\textit{\emph{Color
online: Correlation spreading in long-range spin and boson models
for various values of $\alpha$. (a) Connected spin-spin correlation
function in the LRTI model for the quench $V_{i}=h/2\rightarrow V_{f}=h/10$.
(b) Connected density-density correlation function in the LRBH model
for the quench}}\textit{ $U_{i}=V_{i}=J\rightarrow U_{f}=V_{f}=J/4$
}\textit{\emph{. Results obtained using the t-VMC approach for systems
of $L=400$ sites (for visibility, only a part is shown). The length
unit is the lattice spacing and the time units are $\hbar/h$ for
(a) and $\hbar/J$}} for (b).}
\end{figure*}

In this Rapid Communication, we study the out-of-equilibrium dynamics,
induced by an interaction quench, of homogeneous 1D quantum systems
with long-range algebraic interactions. We consider two different
models, namely the long-range transverse Ising (LRTI) and Bose-Hubbard
(LRBH) models. A quantitative analysis of quasi-locality in these
systems is realized upon studying the earliest times at which information
arrives at some fixed distance. On one hand, we perform \emph{ab-initio}
quantum many-body calculations based on the time-dependent variational
Monte Carlo (t-VMC) approach \cite{carleo2012}. On the other hand,
we provide a unified analytical framework based on quasi-particle
(QP) analysis. Both approaches consistently show that the two systems
behave dramatically different. For $\alpha>2$ the LRTI model shows
ballistic spreading of correlations with exponentially small leaks
in time, hence leading to a strong form of quasi-locality. For $1<\alpha<2$
quasi-locality is still found. However, algebraic leaks in time appear,
which can be traced back to a divergent group velocity at low momenta.
For $\alpha<1$ quasi-locality is instead completely broken. This
effect is traced back to the divergences of both the QP energy and
velocity, which induce infinitely fast oscillations and a response
time scale that goes to zero with the system size. Conversely, for
the LRBH model, we find ballistic spreading of correlations for any
value of $\alpha$, analogous to what found in the short-range Bose-Hubbard
model and in marked contrast with expectations based on the lack of
LR bounds for these systems. This effect is traced back to the fact
that the QP energy remains finite, which cancels non-local contributions
for any value of $\alpha$. All the observed regimes are explained
by the unified QP analysis and shed new light on the microscopic origin
of locality in long-range quantum systems.

\paragraph*{Long-range transverse Ising model.--}

We start with the long-range transverse Ising (LRTI) model, whose
Hamiltonian reads 
\begin{equation}
\mathcal{H}=-h\sum_{i}\sigma_{i}^{x}+\frac{V}{2}\sum_{i\neq j}\dfrac{\sigma_{i}^{z}\sigma_{j}^{z}}{\left|i-j\right|^{\alpha}},\label{eq:ising}
\end{equation}
where $\sigma_{i}^{x},\sigma_{i}^{z}$ are the Pauli matrices, $h$
is the transverse field, $V$ is the strength of the long-range spin
exchange term, and in the following we set $\hbar=1$ for convenience.
Hamiltonian~(\ref{eq:ising}) is the prototype for long-range interacting
quantum systems \cite{tagliacozzo2013,manmana2014,VDW2013}. Moreover,
it is experimentally implemented in cold ion crystals \cite{cirac2004}.
Evidence of the breaking of quasi-locality in information spreading
has been reported for the 1D LRTI model for sufficiently small exponents
$\alpha$ \cite{tagliacozzo2013,manmana2014,porras2014,jurcevic2014},
consistently with the absence of a long-range LR bound for $\alpha<1$.
It was pointed out, however, that a model-dependent form of quasi
locality may occur for specific initial states \cite{manmana2014}
with a complete understanding of the possible scenarios being debated.

Asymptotically reliable results to reveal quasi-locality require sufficiently
long propagation times and sufficiently large systems. This is particularly
crucial to determine precisely the nature of the dynamical regimes.
To achieve this goal, we compute the unitary evolution of the correlation
functions by means of the t-VMC approach \cite{carleo2012,carleo2013}
(see Suppl. Mat.). The latter permits to simulate the dynamics of
correlated quantum systems with an accuracy comparable to tensor-network
methods and proved numerically stable for unprecedented long times
and large sizes. 

In the t-VMC calculations, we use a Jastrow wavefunction with long-range
spin-spin correlations at arbitrary distance \cite{Franjic:1997aa}.
In order to avoid misleading finite size effects, which are usually
strong in these issues \cite{verstraete2014}, periodic boundary conditions
(PBC) are used. For a lattice of size $L$ with PBC, the interaction
potential is taken as the sum of the contributions resulting from
all the periodic images of the finite system. The Fourier components
of the effective interaction potential are then $P\left(k\right)=2\sum_{n=1}^{\infty}\dfrac{\cos\left(kn\right)}{n^{\alpha}}=2\text{Cl}_{\alpha}(k),$
where we have used the Poisson summation formula over the periodic
images, $k$ is an integer multiple of $2\pi/L$, and $\text{Cl}_{\alpha}(k)$
is the Clausen cosine function. In order to have a well-behaved potential
in the thermodynamic limit, we set $P\left(k=0\right)=0$. It is the
equivalent of the standard regularization procedure ensuring charge
neutrality in the presence of electrostatic interactions \cite{ashcroft1976}. 

We consider global quenches of the strength of the long-range interaction,
$V_{i}\rightarrow V_{f}$. The results for the time-connected average
$G_{\text{c}}^{\sigma\sigma}(R,t)=G^{\sigma\sigma}(R,t)-G^{\sigma\sigma}(R,0)$
of the spin-spin correlation function $G^{\sigma\sigma}(R)=\left\langle \sigma_{i}^{z}\sigma_{i+R}^{z}\right\rangle $
are shown in Fig.~\ref{fig:tvcm-corr}~(a). We find three qualitatively
different regimes. For $\alpha<1$, Fig.~\ref{fig:tvcm-corr}~(a1),
the propagation of correlations takes place on extremely short time
scales and no cone-like structure emerges. This is the signature of
an efficient microscopic mechanism leading to the breaking of locality
in the system. For $\alpha>2$, Fig.~\ref{fig:tvcm-corr}~(a3),
a correlation cone with a well-determined velocity $v$ clearly emerges.
It is marked by a strong suppression of leaks in the region defined
by $R/t>v$ and space-time oscillations in the region $R/t<v$. In
the intermediate regime where $1<\alpha<2$, Fig.~\ref{fig:tvcm-corr}~(a2),
a correlation cone-like structure is still visible but prominent non-local
leaks are also appearing.

\begin{figure}[t]
\includegraphics[clip,width=1\columnwidth]{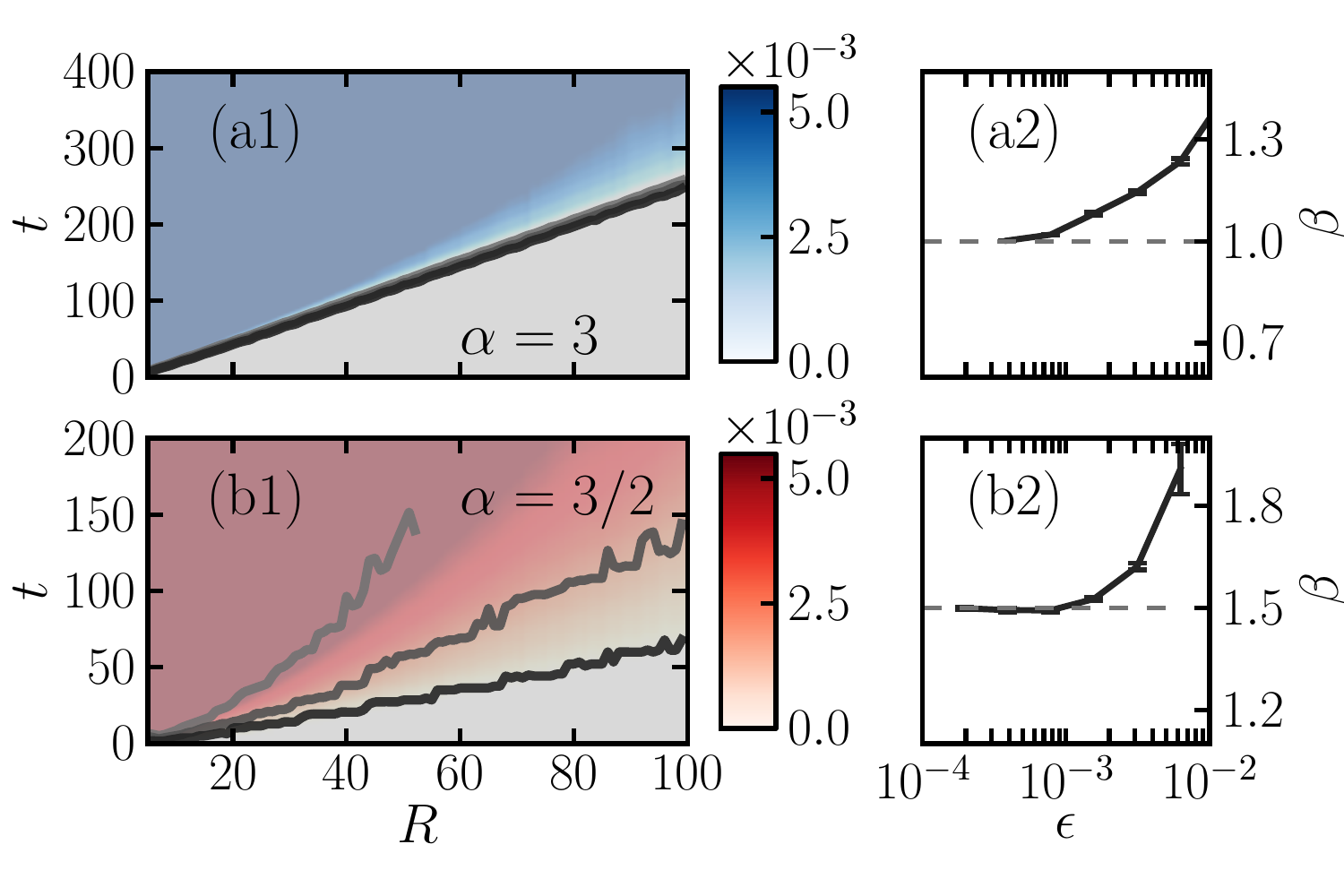}\caption{\textit{\footnotesize \label{fig:epslan}}\textit{\emph{(Color online)
Panels (a1,b1): Time-integrated spin-spin correlation functions $\bar{G}_{c}^{\sigma\sigma}(R,t)$
for the same data as in Fig. \ref{fig:tvcm-corr}}}~\textit{\emph{(a).
Superimposed lines show the activation time $t^{\star}(\epsilon)$
(see text), for $\epsilon$ ranging from $2\times10^{-2}$ (lighter
lines) down to $5\times10^{-3}$ (darker lines). Panels (a2,b2): Behavior
of the fitted exponent $\beta$, computed within linear spin-wave
theory, as a function of the cutoff parameter $\epsilon$ and for
the system size $L=2^{14}$.}}\textit{ }\textit{\emph{The length unit
is the lattice spacing and the time unit is $\hbar/h$}}.}
\end{figure}

In order to quantify more precisely the time-dependence of the leaks
in the quasi-local regimes we study the time-integrated absolute value
of the correlation function $\bar{G}_{c}^{\sigma\sigma}(R,t)=\frac{1}{t}\int_{0}^{t}dt^{\prime}\left|G_{c}^{\sigma\sigma}(R,t^{\prime})\right|$.
While it retains all the features of the signal propagation, it is
less sensitive to time oscillations. In Fig.~\ref{fig:epslan} (a1)
we show the behavior of $\bar{G}_{c}^{\sigma\sigma}(R,t)$ for $\alpha=3$.
It clearly shows the sharp boundary of a ballistic cone. This is further
assessed introducing a small cutoff $\epsilon$ and computing the
first propagation time $t^{\star}(R)$ such that $\bar{G}_{c}^{\sigma\sigma}(R,t^{\star})>\epsilon$.
The result is almost independent of $\epsilon$ and we find the scaling
$vt^{\star}=R^{\beta}$, with finite $v$ and $\beta\simeq1$ to very
good precision and up to large system sizes and long propagation times.
The presence of a ballistic spreading, with exponentially suppressed
leaks in time, is a stronger realization of locality than what expected
from the looser long-range LR bound, which instead allows for polynomially
suppressed leaks in time %
\footnote{We focus here exclusively on the time-dependence of the leaks, and
we find regimes with exponentially suppressed leaks in time. Notice
that the spatial dependence of the correlation function, even for
short-range Hamiltonians, can instead exhibit algebraic behavior,
associated to quasi-long-range order in 1D. %
}. For $1<\alpha<2$ the same analysis of the leaks, shown in Fig.~\ref{fig:epslan}
(b1), reveals instead that polynomial leaks in time appear with an
exponent $\beta\simeq\alpha$, and a velocity $v$ that vanishes with
$\epsilon$. This is compatible with the long-range LR bound~\cite{Hastings2010}.
Remarkably, the regimes we find here for a global quench are the same
qualitative regimes that have been identified for a local quench in
the LRTI model in Ref. \cite{tagliacozzo2013}. 

\paragraph*{Long-range Bose-Hubbard model.--}

We now turn to the long-range Bose-Hubbard (LRBH) model, which describes
interacting spinless bosons in a periodic potential with nearest-neighbor
tunneling and long-range two-body interactions. The Hamiltonian reads
\begin{equation}
\mathcal{H}=-J\sum_{\langle i,j\rangle}\left(b_{i}^{\dagger}b_{j}+h.c.\right)+\frac{U}{2}\sum_{i}n_{i}\left(n_{i}-1\right)+\frac{V}{2}\sum_{i\neq j}\frac{n_{i}n_{j}}{|i-j|^{\alpha}},\label{LRBH}
\end{equation}
where $b_{i}(b_{i}^{\dagger})$ destroys (creates) a boson on site
$i$, $n_{i}=b_{i}^{\dagger}b_{i}$ is the particle number operator,
$J$ is the tunneling amplitude, $U$ is the on-site interaction energy,
$V$ is the strength of the interaction potential, and we set $\hbar=1$
for convenience. The short-range ($V=0$) case is now routinely realized
in ultracold-atom experiments and the long-range ($V\neq0$) case
applies to polar molecules \cite{Buchler:2009aa}, magnetic \cite{menotti2008}
and Rydberg atoms \cite{pohl2014}.

We perform t-VMC calculations using a Jastrow wavefunction incorporating
density-density correlations at arbitrary large distances \citep{PhysRevLett.99.056402}
(see Suppl. Mat.). For simplicity we choose $U=V$, fix the density
at half filling ($n=\frac{1}{2}$), and consider the connected density-density
correlation function $G_{\text{c}}^{nn}\left(R\right)=\left\langle n_{i}n_{i+R}\right\rangle -n^{2}$.
We study global quenches in the interaction strength $V_{i}\rightarrow V_{f}$.
The results for various values of the exponent $\alpha$ are shown
in Fig.~\ref{fig:tvcm-corr}~(b). Surprisingly, we find here that
the LRBH model exhibits the same qualitative behavior for all values
of $\alpha$, in marked contrast with the LRTI model. Within numerical
precision, we always find a purely ballistic cone spreading of correlations
at some velocity $v$. The long-range LR bound is therefore never
saturated and, for every value of the exponent $\alpha$, the spreading
is qualitatively identical to the short-range case. Hence, in the
LRBH model quasi-locality appears to be strongly protected even for
very long-range interactions. This is further confirmed by a precise
analysis of the leaks, along the same lines as for the LRTI model.
It always yields a scaling of the form $vt^{\star}=R^{\beta}$ with
$\beta=1$ and the signal is exponentially suppressed for times out
of the locality cone, i.e. when $t<R/v$.

\paragraph*{Quasi-particle analysis.--}

The radically different behaviors of the LRTI and LRBH models are
particularly striking because they share the same class of long-range
interactions and are therefore subjected to the same universal long-range
LR bounds \cite{Hastings2010}. In order to understand the different
behaviors of the two models, at a miscroscopic level, we use a general
QP approach. The latter has a broad range of applications, e.g.\ universal
conformal theories~\cite{calabrese2006}, spin systems~\cite{tagliacozzo2013},
superfluids~\cite{natu2013}, and Mott insulators~\cite{cheneau2012}.
A generic time-dependent two-body, connected correlation function
in a translation invariant model with well-defined QP excitations
can be written as 
\begin{multline}
G_{\text{c}}\left(R,t\right)=\int_{-\pi}^{\pi}\frac{dk}{2\pi}\mathcal{F}\left(k\right)\Bigl\{\cos\left(kR\right)+\\
\left.-\frac{1}{2}\left[\cos\left(kR-2E_{k}^{f}t\right)+\cos\left(kR+2E_{k}^{f}t\right)\right]\right\} ,\label{eq:GrGeneral}
\end{multline}
where $E_{k}^{f}$ is the $k$-momentum QP energy of the post-quench
Hamiltonian and $\mathcal{F}\left(k\right)$ is the weight associated
to each QP. This general form states that the collective excitations
of the system are coherent superpositions of pairs carrying excitations
of momentum $k$ and traveling in opposite directions. Whereas $E_{k}^{f}$
depends only on the post-quench Hamiltonian, in general $\mathbf{\mathcal{F}}\left(k\right)$
instead depends both on the pre- and post-quench Hamiltonians %
\footnote{The validity of the QP picture in this context is corroborated by
the good quantitative agreement between the QP dynamics given by Eq.
\eqref{eq:GrGeneral} and the t-VMC results (see Suppl. Mat.).%
}.

In the LRTI model and in the regime of large transverse field ($h\gg V$),
considered in the t-VMC calculations, we can apply linear spin wave
theory \citep{PhysRevA.72.063407}. The QP energy and weight read,
respectively, $E_{k}^{f}=2\sqrt{h[h+VP(k)]}$ and $\mathcal{F}^{\sigma\sigma}\left(k\right)=\dfrac{2P(k)(V_{i}-V_{f})}{E_{k}^{i}\left[1+P(k)V_{f}/h\right]},$
where $P(k)$ are the Fourier components of the interaction potential
(see Suppl. Mat.). Let us then analyze the outcome of Eq.~(\ref{eq:GrGeneral}).

The ballistic behavior observed for $\alpha>2$ can be understood
from stationary phase analysis. Along the line $R=vt$, it yields
the dominant contribution 
\begin{eqnarray}
G_{\text{c}}^{\sigma\sigma}(R,t) & \simeq & \frac{\mathcal{F}(k^{\ast})}{4\sqrt{\pi\left|\partial_{k}^{2}E_{k^{\ast}}^{f}\right|t}}\cos\left(k^{\ast}R-2E_{k^{\ast}}^{f}t+\phi\right),\label{eq:statphase}
\end{eqnarray}
where $k^{\star}$ is solution of $v=2v_{g}(k^{\star})$, $v_{g}$
is the group velocity, and $\phi$ is a time-independent phase. For
$\alpha>2$, the group velocity is bounded and we find a ballistic
cone spreading of correlations with a velocity that is given by twice
the maximum group velocity. We have checked that it agrees well with
the numerics where a cone propagation is also found. A quantitative
analysis of the leaks within the spin-wave approach confirms that
they are exponentially suppressed in time. Moreover, the exponent
$\beta$ is found to smoothly approach unity upon reducing the cutoff
parameter $\epsilon$ {[}see Fig.~\ref{fig:epslan} (a2){]}, in agreement
with the t-VMC results.
\begin{figure}[t]
\includegraphics[clip,width=1.05\columnwidth]{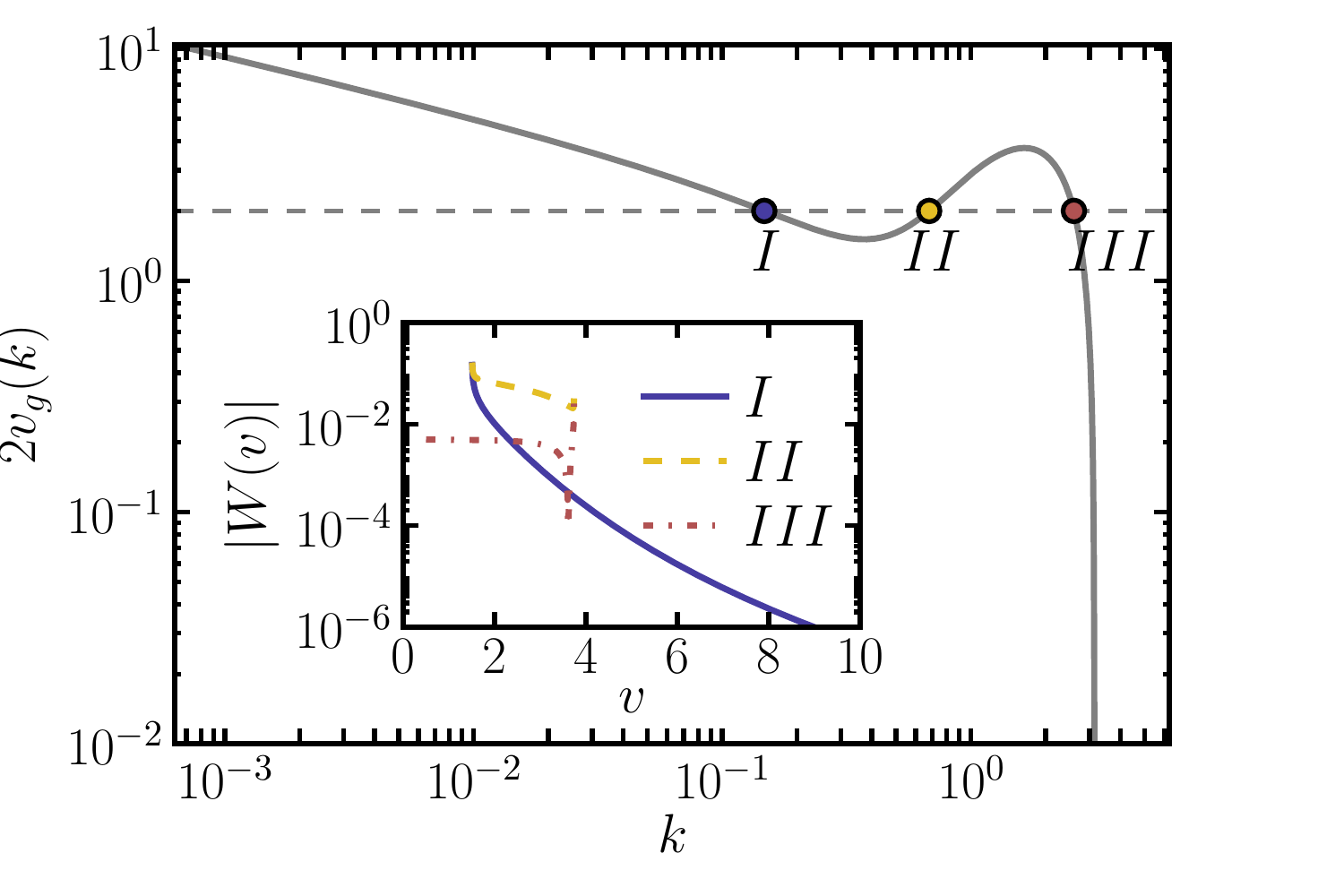}\caption{\textit{\footnotesize \label{fig:wsf} }\textit{\emph{(Color online)
QP analysis of the group velocity for the LRBH model with exponent
$\alpha=1/2$, density $n=1/2$ and $U=V=J/4$. In the inset, the
stationary phase weights corresponding to the three solution branches
of the equation $2v_{g}(k)=v$ are shown.}} \textit{\emph{Wave-vectors
are in units of the inverse lattice spacing and velocities are in
units of the hopping amplitude $J$}}.}
\end{figure}

For $1<\alpha<2$ the group velocity is instead unbounded, since it
exhibits the infrared divergence $v_{g}(k)\propto k^{-\left|2-\alpha\right|}$.
Hence the correlation front is no longer given by a well-determined
velocity but by the coherent superposition of infinitely fast modes
at low momenta. More precisely, the behavior of the leaks can be found
from the asymptotic $R\rightarrow\infty$ expansion of Eq. \eqref{eq:GrGeneral},
retaining only the contributions of the divergent velocities. For
$\alpha=3/2$, it can be computed exactly, and yields $G_{\text{c}}^{\sigma\sigma}(R\rightarrow\infty,t)\simeq F(t)\times t/R^{\alpha}$,
where $F(t)$ is a bounded function of time. This scaling agrees with
and explains the t-VMC results. It is further confirmed by the analysis
of the leaks within the spin-wave approach. The exponent $\beta$
is found to smoothly approach the interaction exponent $\alpha$ upon
reducing the cutoff parameter $\epsilon$ {[}see Fig.~\ref{fig:epslan}
(b2){]}. The same was found for other values of $\alpha$ between
$1$ and $2$.

For $\alpha<1$, both the QP energy and the group velocity diverge
for $k\rightarrow0$, respectively, as $E_{k}^{f}=e_{0}k^{-\left|\frac{1-\alpha}{2}\right|}$,
$v_{g}(k)\propto k^{-\left|\frac{3-\alpha}{2}\right|}$ and $e_{0}=2\sqrt{h_{f}V_{f}}$.
In particular, it is the energy divergence which sets the breaking
of quasi-locality in this case. The latter gives rise to a rapidly
oscillating factor of the form $\cos\left(e_{0}t/k^{\frac{1-\alpha}{2}}\right)$
in Eq.~\eqref{eq:GrGeneral}, which leads to a non-analytic $t=0$
delta-kick in the thermodynamic limit. More precisely, an exact asymptotic
expansion of the correlation function \eqref{eq:GrGeneral} can be
derived in the limit of small propagation times $t$ and large distances
$R$. Keeping the relevant small quasi-momenta, it yields 
\begin{equation}
G_{\text{c}}^{\sigma\sigma}(R,t\rightarrow0)\simeq\lim_{L\rightarrow\infty}A\frac{\sin\left(L^{\frac{1-\alpha}{2}}e_{0}t\right)}{e_{0}t}\frac{\cos\left(R/L\right)}{R^{2-\alpha}}+B\frac{\left(e_{0}t\right)^{2}}{R^{\frac{1+\alpha}{2}}},\label{eq:asympsmallt}
\end{equation}
where $A$ and $B$ are finite numerical constants, which depend on
the microscopic parameters of the model. A remarkable consequence
of this expression is that the first term yields an instantaneous
contribution to the signal, on a time scale $e_{0}\tau=1/L^{\frac{(1-\alpha)}{2}}$,
independent of the distance $R$ and with an exponent set by the divergence
of the QP energy. This implies that the system reacts on a time scale
inversely proportional to the system size, yielding an efficient mechanism
for the breaking of locality.

An analogous microscopic analysis can be carried out for the LRBH
model in the weakly interacting superfluid regime, considered in the
t-VMC calculations. In this regime the quantities $E_{k}^{f}$ and
$\mathcal{F}\left(k\right)$ are found by Bogoliubov analysis \citep{natu2013},
which yields $E_{k}^{f}=\sqrt{\epsilon_{k}\left[\epsilon_{k}+2n\left(U_{f}+V_{f}P\left(k\right)\right)\right]}$
and $\mathcal{F}^{nn}\left(k\right)=n^{2}\frac{\left[\left(U_{f}-U_{i}\right)+\left(V_{f}-V_{i}\right)P\left(k\right)\right]\epsilon_{k}}{\left[\epsilon_{k}+2n\left(U_{f}+V_{f}P\left(k\right)\right)\right]E_{k}^{i}}$,
where $\epsilon_{k}=4J\sin^{2}\left(k/2\right)$ is the free-particle
lattice dispersion and $n$ is the particle density (see Suppl. Mat.).

For $\alpha>1$, the origin of the observed ballistic behavior is
traced back to the fact that QP velocities are bounded. The propagation
of correlations is dominated by the stationary-phase points of Eq.
\eqref{eq:statphase} and the correlation cone velocity is given by
twice the maximum group velocity. 

For $\alpha<1$, the group velocity diverges as $v_{g}(k\rightarrow0)\propto k^{-\left|\frac{1-\alpha}{2}\right|}$,
whereas the QP energy is always finite. Hence, at variance with the
LRTI model, the correlation function does not exhibit any instantaneous
kick at $t=0$ such as that of Eq. \eqref{eq:asympsmallt} and quasi-locality
is preserved. Moreover, although this case is formally analogous to
the intermediate regime, with polynomial leaks in time, found for
the LRTI model, a purely ballistic spreading is instead found within
numerical precision in the LRBH model. To understand this, let us
come back to the stationary-phase approach of Eq. \eqref{eq:statphase}.
Due to the specific form of the group-velocity dispersion in the LRBH
model, shown in Fig. \ref{fig:wsf}, the equation $v=2v_{g}(k^{\star})$
has up to three separate solutions for a given velocity $v$. The
correlation function is thus dominated by three contributions ($I$,
$II$, and $III$) of the form of Eq. \eqref{eq:statphase}. The behavior
of the corresponding weights, $W(v)=\mathcal{F}^{nn}\left(k^{\star}\right)/\sqrt{\left|\partial_{k}^{2}E^{f}(k^{\star})/J\right|}$,
along the three branches is shown in the inset of Fig. \ref{fig:wsf}.
The largest weights, corresponding to the velocities dominating the
propagation, belong to the regular branches ($II$ and $III$). The
latter extend up to a certain maximum velocity $v_{\text{max}}$,
which effectively sets the correlation cone velocity $v_{\text{c}}$.
The infinitely fast modes, corresponding instead to branch $I$, have
a weight which is polynomially suppressed at large velocities. The
exponent for the algebraic decay can be derived from the known small
$k$ behavior of $P(k)$, which leads to $W(v\rightarrow\infty)\propto v^{-\left|\frac{9-3\alpha}{2(1-\alpha)}\right|}$.
For $v\simeq v_{\text{max}}$, the weights of these modes are already
few orders of magnitude smaller than the quasi-local, ballistic modes.
This separation of scales is responsible of the effective suppression
of the infinitely fast non-local modes in the LRBH model. Notice that
in the LRTI model the irregular branch of the infinitely fast modes
is never protected by a finite-velocity branch. This is a consequence
of the monotonic behavior of the QP group velocity in the spin model,
yielding a single branch of solutions for the stationary phase equation.

\paragraph*{Discussion.--}

In summary, the radically different behaviors of the LRTI and LRBH
models, found both in t-VMC and QP analysis, show that specific microscopic
properties of the system, not accounted for in universal LR-like bounds,
play a major role in quasi-locality. This result sheds new light on
the dynamics of long-range quantum systems. Yet many questions remain
open and are worth being investigated in the future. For instance,
it is expected, on the basis of universal bounds, that the critical
exponents for the breaking of locality depend on the system dimension~\cite{Hastings2010}.
It would thus be of utmost interest to study the counterparts of the
effects discussed here in dimension higher than one, which could be
done by a straightforward application of the present approach~\cite{carleo2013}.
Moreover, it would be interesting to study the same LRTI and LRBH
models in a regime of stronger interactions, where an extension of
the t-VMC and QP analysis taking into account relevant excitations
can be developed. In this regime, the LRBH model maps onto an effective
spin model and might therefore exhibit radically different properties
than those found in this work. 
\begin{acknowledgments}
We acknowledge discussions with M. Cheneau, M. Fagotti, M. Holtzmann,
D. Porras, and L. Tagliacozzo. This research was supported by the
European Research Council (FP7/2007-2013 Grant Agreement No. 256294),
Marie Curie IEF (FP7/2007- 2013 - Grant Agreement No. 327143), FET-Proactive
QUIC (H2020 grant No. 641122). It was performed using HPC resources
from GENCI-CCRT/CINES (Grant c2015056853). Use of the computing facility
cluster GMPCS of the LUMAT federation (FR LUMAT 2764) is also acknowledged. 
\end{acknowledgments}
%

\end{document}